\documentclass{article}
\usepackage{graphicx}

\begin{document}

\title{The impact of celestial pole offset modelling on VLBI UT1 Intensive results}
\author{Zinovy Malkin \\ Pulkovo Observatory, Pulkovskoe~Ch. 65, St.~Petersburg 196140, Russia \\malkin@gao.spb.ru}
\date{~}
\maketitle

\begin{abstract}
Very Long Baseline Interferometry (VLBI) Intensive sessions are scheduled to provide operational Universal Time (UT1) determinations with low latency.
UT1 estimates obtained from these observations heavily depend on the model of the celestial pole motion used during data processing.
However, even the most accurate precession-nutation model, IAU~2000/2006, is not accurate enough to realize the full potential of VLBI observations.
To achieve the highest possible accuracy in UT1 estimates, a celestial pole offset (CPO), which is the difference between the actual and modelled
pre\-ces\-si\-on-nu\-ta\-ti\-on angles, should be applied.
Three CPO models are currently available for users.
In this paper, these models have been tested and the differences between UT1 estimates obtained with those models are investigated.
It has been shown that neglecting CPO modelling during VLBI UT1 Intensive processing causes systematic errors in UT1 series of up to 20~$\mu$as.
It has been also found that using different CPO models causes the differences in UT1 estimates reaching 10~$\mu$as.
Obtained results are applicable to the satellite data processing as well.
\end{abstract}

\section{Introduction}

Very long baseline interferometry (VLBI) is the principal method for determination of the Universal Time (UT1) nowadays.
The most accurate UT1 estimates can be obtained, simultaneously with other Earth rotation parameters (EOP): terrestrial and celestial pole
coordinates, from 24-hour VLBI sessions carried out on global networks of several, usually 6 to 8, stations.
Their precision and accuracy depends on several factors, such as the number of stations, network geometry, and registration mode (see e.g. Malkin 2009).
However, results of these observations are normally available only in 8 to 15 days after observations.
Such a delay of the results causes problems in many real-time applications and influences UT1 prediction accuracy (Luzum \& Nothnagel 2010).

To get more timely UT1 estimates, special observing programs are conducted in the framework of the International VLBI Service for Geodesy and
Astrometry (IVS) activity (Schl\"uter \& Behrend 2007).
These sessions, called Intensives, are carried out on one or two baselines, have 1-hour duration, employ electronic data transfer (e-VLBI),
and hence provide rapid turnaround time, usually 0.5 to 2 days.
However, the Intensives do not allow us to obtain all the EOP with reasonable accuracy, and only UT1 can be effectively estimated.
For this reason, reliable terrestrial and celestial pole coordinates are needed for data processing of the Intensives.
In this paper we concentrate on the latter aspect.

The PN model errors substantially influence UT1 estimates (Robertson et al. 1985, Titov 2000, Nothnagel \& Schnell 2008).
In particular, the authors of the latter paper have shown that an error in PN of 1~mas can induce an error in UT1 of up to 30~$\mu$as,
and this error is systematic in its behavior.
Since the 24-hour VLBI CPO estimates are yet not available during operational processing of the Intensives, the CPO model should also include prediction
to provide operational Intensives data processing.

The most accurate celestial pole (CP) coordinates can be computed as a sum of the reference precession-nutation (PN) model, currently IAU~2000/2006,
and the observed CP offset (CPO).
Three CPO models suitable for operational processing of the Intensives are publicly available.
They are computed at the U.~S. Naval Observatory (IERS Annual Report 2007), Paris Observatory (Petit \& Luzum 2010),
and Pulkovo Observatory (Malkin 2007).
They are compared in Section~2.
All three models are empirical in their nature, and are evaluated by approximation and extrapolation of an observed VLBI CPO series.
The authors of these models use different algorithms and reference CPO series, which results in systematic differences between models,
and consequently between Intensives UT1 estimates.

In this paper, we investigated this differences with emphasis on their systematic part, which is of main concern for the users.
For this purpose we performed test processing of the Intensives observations obtained during 2007.0--2010.5.
The results are presented and discussed in Section~3.
On this basis, we discuss the choice of the best CPO model to be recommended for using in the Intensives data processing.

\section{CPO models}

CPO consists of several terms, and can be described by the following formula.
\begin{equation}
CPO = FCN + \sum T_i + \sum P_j \,,
\label{eq:cpo-fcn}
\end{equation}
where $T_i$ are trend components caused e.g. by errors in precession or very-low-frequency nutation terms, $P_j$ are (quasi)harmonic terms
caused e.g. by errors in nutation model or geophysical processes.
FCN stands for Free Core Nutation, it is a quasi-harmonic CP motion with an amplitude of about 150 to 200~$\mu$as nowadays.
The sum of the trend components is in our days of the same order as FCN, and is progressively growing.

All three models introduced above are provided as a regularly updated time series of the CPO components $dX$ and $dY$ with prediction for several months.
These models are as follows:
\begin{description}
\item{\it NEOS series\footnote{http://maia.usno.navy.mil/}}
  computed at the U. S. Naval Observatory by spline smoothing of several selected input VLBI CPO series (IERS Annual Report 2007).
  It is an official product of the International Earth Rotation and Reference Systems Service (IERS).
\item{\it SL series\footnote{http://syrte.obspm.fr/~lambert/fcn/}} (S. Lambert)
  computed at the Paris Observatory by least-squares fit of the circular FCN term with period of 430.2 days and a bias over the running 2-yr intervals.
  IERS C04 EOP series is used as input.
  The bias is not included in the model.
  The SL model is recommended by the IERS Conventions (2010) as a replacement for CPO (Petit \& Luzum 2010).
\item{\it ZM2 series\footnote{http://www.gao.spb.ru/english/as/persac/index.htm}}
  computed at the Pulkovo Observatory by exponential smoothing of the IVS combined CPO series (Malkin 2007).
\end{description}

The IVS combined EOP series is an official IVS product computed by combining datum-free normal equations of up to six IVS analysis centers
(B\"ockmann 2010).
It includes estimates of Pole coordinates with their rates, Universal Time, Length of Day, and CPO for every 24-hour VLBI session suitable for accurate
EOP determination, two to three sessions a week in average.
This series, along with supplement information, is available at the IVS Analysis Coordinator Web site\footnote{http://vlbi.geod.uni-bonn.de/IVS-AC/},
and is supposed to be the most accurate source to our knowledge of the CP motion.

\begin{figure}
\centering
\includegraphics[clip,width=0.5\hsize]{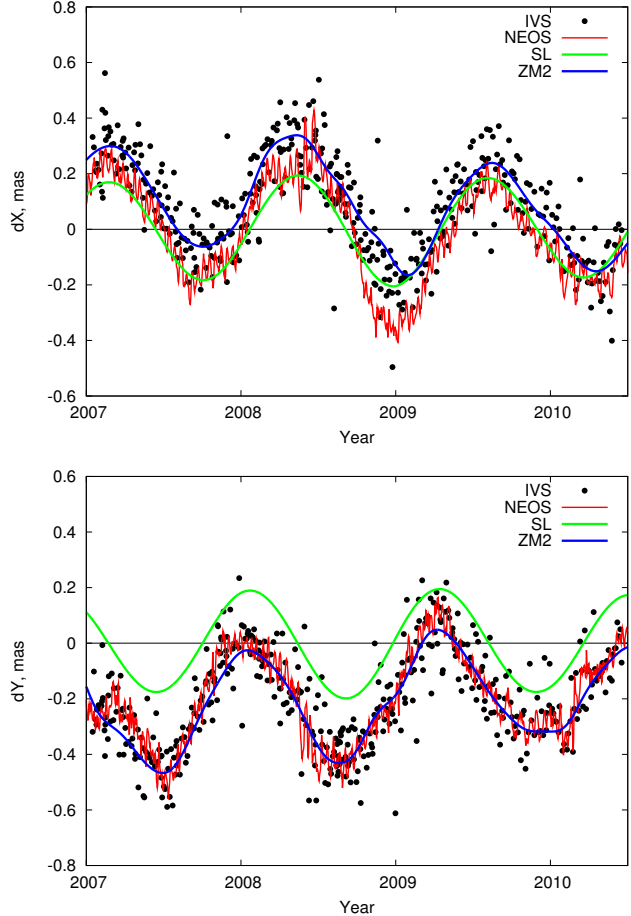}
\caption{Comparison of three CPO models with the IVS CPO combines series. Unit: mas.}
\label{fig:cpo_ivs}
\end{figure}

\begin{figure*}
\centering
\includegraphics[clip,width=0.9\hsize]{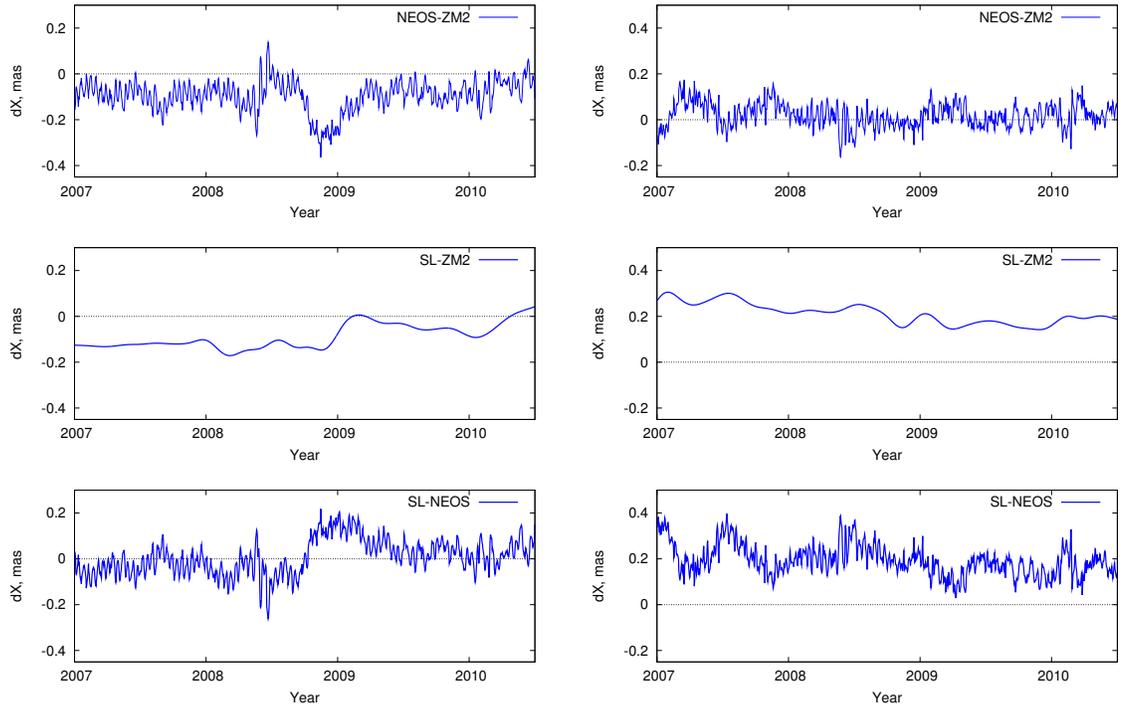}
\caption{Differences between CPO models. Unit: mas.}
\label{fig:cpo_dif}
\end{figure*}

Figure~\ref{fig:cpo_ivs} shows a comparison of the CPO models with the IVS CPO series, and Fig.~\ref{fig:cpo_dif} presents the differences between
the three models.
One can see by visual inspection that the ZM2 model represent IVS data better than others, just because of how it is evaluated.
The NEOS model is close to ZM2, but seems to have extra high-frequency terms, which most probably comes from the incoming data noise.
A bias between ZM2 and NEOS models seen in some periods evidently comes from the difference between the IVS and NEOS combined CPO series.
The SL model is centered around the time axis and has a significant bias with respect to the IVS series, which can be expected from the method
used for its computation, as described above.
This bias corresponds to two last terms of (\ref{eq:cpo-fcn}) and is variable with time.
Table~\ref{tab:cpo-ivs} shows the rms difference between each of the CPO models and the IVS combined CPO series at the interval 2007--2010.5,
(that is, the rms difference of the series shown in Fig.~\ref{fig:cpo_ivs}).

Detailed analysis of the differences between CPO models and their origins is beyond the scope of this paper, although it is an interesting task
which is worth a separate study.
However, our comparison clearly shows that significant differences up to several hundred~$\mu$as between the CPO models do exist, which can induce
differences between the UT1 estimates computed using various models.

\begin{table}
\centering
\caption{The rms difference between each of the CPO models and the IVS combined CPO series. Unit: $\mu$as.}
\label{tab:cpo-ivs}
\tabcolsep=13.3pt
\begin{tabular}{cc|cc|cc}
\hline
\multicolumn{2}{c}{NEOS} & \multicolumn{2}{|c|}{SL} & \multicolumn{2}{c}{ZM2} \\
\hline
$dX$ & $dY$ & $dX$ & $dY$ & $dX$ & $dY$ \\
\hline
 159 &  ~83 &  129 &  217 &  ~70 &  ~79 \\
\hline
\end{tabular}
\end{table}

\section{Data analysis}

\begin{figure}
\centering
\includegraphics[clip,width=0.4\hsize]{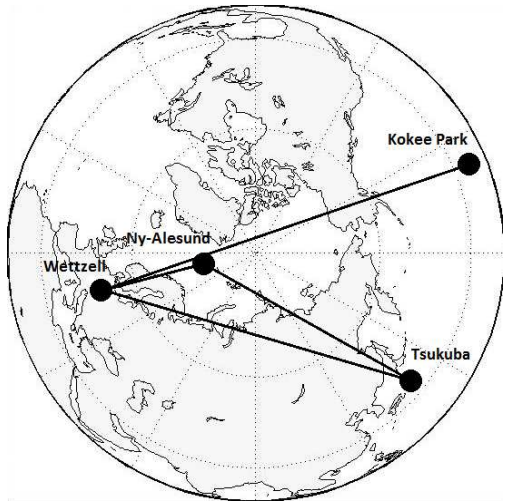}
\caption{Stations and baselines participating in the analyzed observing programs.}
\label{fig:map}
\end{figure}

This study is based on the processing of VLBI data collected from three main Intensives programs from 2007.0 till 2010.5.
The observations from the following observing programs were used (see also Fig.~\ref{fig:map}):
\begin{description}
\item{\it Int1:} Kokee Park---Wettzell baseline, 803 sessions;
\item{\it Int2:} Tsukuba---Wettzell baseline, 462 sessions;
\item{\it Int3:} Ny-{\AA}lesund---Tsukuba---Wettzell baselines, 96 sessions.
\end{description}

\begin{figure*}
\centering
\includegraphics[clip,width=\hsize]{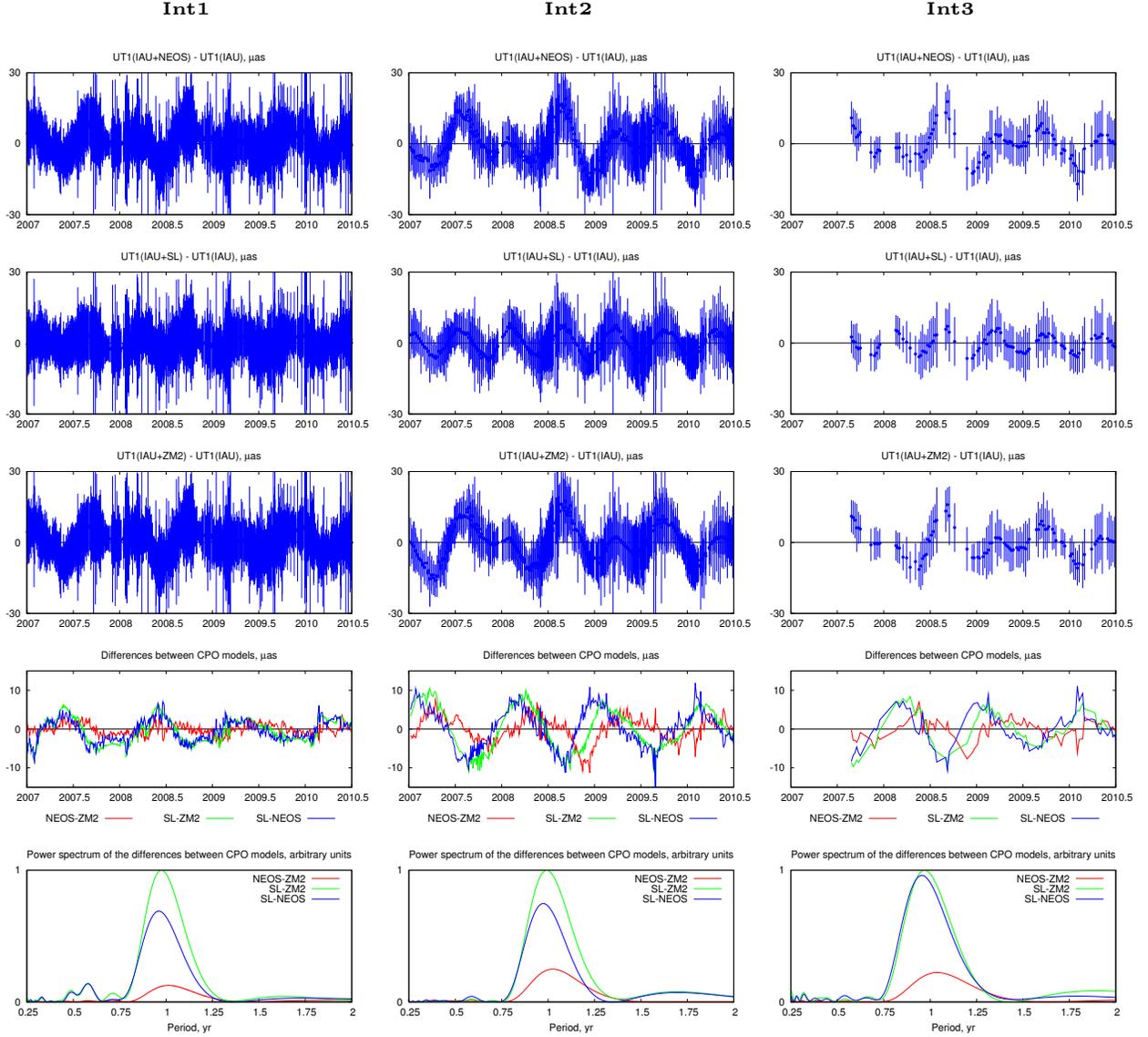}
\caption{The impact of CPO model on UT1 Intensives result. Three columns correspond to three Intensives observing programs.
Three upper plots display the differences between UT1 estimates obtained with one of the given CPO models (from top to bottom: NEOS, SL, ZM2)
and the UT1 estimates obtained without CPO correction, vertical bars show 1$\sigma$ uncertainty. In the 4th raw, the differences between
the UT1 estimates obtained with the three CPO models are plotted, unit: $\mu$as. The lower plots show the spectra of these differences in
arbitrary units.}
\label{fig:dif}
\end{figure*}

All the observations were processed in a uniform way with the only difference in the CPO model used.
Source positions were taken from the ICRF2 (the second realization of the International Celestial Reference Frame) catalog (Ma et al. 2009).
The computations were made in four variants: with one of the three CPO models listed above and without CPO contribution
(in other words, with zero CPO model), and four corresponding UT1 time series were obtained and compared.
Results of this computations and comparison of UT1 estimates are depicted in Fig.~\ref{fig:dif}.

In Table~\ref{tab:error_bias}, a comparison of the formal errors and mutual biases are presented.
The formal errors are those computed during session data processing, and the bias is computed as the average value over the whole 3.5-year period
with respect to the UT1 time series computed without CPO correction.
The results show that the CPO modelling has practically no influence on the UT1 uncertainty, and the relative bias.

\begin{table}
\centering
\caption{Median formal error of the UT1 estimates and relative bias of the UT1 time series computed with different CPO models. Unit: $\mu$as.}
\label{tab:error_bias}
\tabcolsep=7.4pt
\begin{tabular}{l|ccc|ccc}
\hline
Model    & \multicolumn{3}{c|}{Formal error} & \multicolumn{3}{|c}{Bias} \\
\cline{2-7}
         & Int1 & Int2 & Int3 & Int1   & Int2   & Int3   \\
\hline
Zero CPO & 10.3 &  9.1 &  8.0 & $   0$ & $   0$ & $   0$  \\
NEOS     & 10.3 &  9.0 &  8.0 & $-0.7$ & $+0.1$ & $+0.2$  \\
SL       & 10.4 &  9.0 &  8.0 & $-0.1$ & $+0.1$ & $+0.3$  \\
ZM2      & 10.3 &  9.0 &  8.0 & $-0.7$ & $-0.2$ & $-0.1$  \\
\hline
\end{tabular}
\end{table}

However, one can see that neglecting CPO modelling causes time-dependent systematic errors with amplitudes of up to 20~$\mu$as.
As previously noticed by Nothnagel \& Schnell (2008), the effect is larger for Int2 sessions as compared with Int1.
It should be mentioned that differences between UT1 estimates obtained above may also depend on other factors like observation scheduling as it has been
shown in previous investigations of the Intensives results (Hefty \& Gontier 1997, Titov 2000).
This is also confirmed by the results presented in Fig.~\ref{fig:dif} and Table~\ref{tab:annual_term} below which show different influence of CPO
modelling on UT1 time series obtained in the framework of different observing programs.
For this reason, one should be careful about too strict comparison of those differences with the differences in CPO models.
On the other hand, the differences between the UT1 time series obtained with different CPO models depend only on this factor, and allow us to make
a conclusion on its impact on the Intensives results, which is the primary goal of this paper.

Further comparison of the UT1 time series revealed more complicated impact of the CPO modelling on the systematic behavior of UT1 estimates,
as can be seen in Fig.~\ref{fig:dif}, 4th raw, where the systematic differences between UT1 estimates obtained with different CPO models with amplitude
of up to 10~$\mu$as are clearly present.
Since these differences look like (quasi-)periodic signal, we performed spectral analysis (the lower raw) which showed only one statistically
significant harmonic term with period of about one year.
The amplitude and phase of the annual term computed by weighted least-squares adjustment of UT1 differences are presented in Table~\ref{tab:annual_term}.
The weights were computed from the formal errors of UT1 estimates.
The phase is given at the epoch of minimum variance, the same for each observing program.
These results show that the impact of CPO modelling on Int2 results is more than twice larger than that for Int1.
Addition of a third station to Int2, i.e. using the Int3 network, somewhat reduces the amplitude of the annual term, but does not principally change the result.
A comparison of results presented in Table~\ref{tab:annual_term} confirms that the systematic differences between UT1 time series computed with two
full CPO models, NEOS and ZM2, are relatively small, whereas differences with UT1 time series computed with the FCN/SL model are about twice larger.

\begin{table}
\centering
\caption{The amplitude and phase of the annual term in differences between UT1 estimates computed with different CPO models.}
\label{tab:annual_term}
\begin{tabular}{l|ccc|ccc}
\hline
Models      & \multicolumn{3}{c|}{Amplitude, $\mu$as} & \multicolumn{3}{|c}{Phase, deg} \\
\cline{2-7}
            & Int1     & Int2     & Int3     & Int1     & Int2     & Int3     \\
\hline
NEOS -- ZM2 &    ~~1.3 &    ~~3.8 &    ~~2.8 &      175 &      235 &      341 \\
            & $\pm$0.1 & $\pm$0.1 & $\pm$0.2 &   $\pm$2 &   $\pm$2 &   $\pm$5 \\[1ex]
SL -- ZM2   &    ~~3.5 &    ~~7.5 &    ~~6.1 &      220 &      292 &      ~50 \\
            & $\pm$0.1 & $\pm$0.1 & $\pm$0.3 &   $\pm$1 &   $\pm$1 &   $\pm$3 \\[1ex]
SL -- NEOS  &    ~~2.7 &    ~~6.3 &    ~~5.8 &      239 &      322 &      ~77 \\
            & $\pm$0.1 & $\pm$0.2 & $\pm$0.3 &   $\pm$2 &   $\pm$1 &   $\pm$3 \\
\hline
\end{tabular}
\end{table}

It is also interesting to notice that the differences between UT1 series NEOS--ZM2 and SL--NEOS are more noisy than that for SL--ZM2.
This may indicate that an extra noise component in the NEOS series propagates to UT1 estimates too.

\section{Conclusions}

In this paper, we investigated and compared the impact of three CPO models on UT1 estimates obtained from VLBI Intensives.
The NEOS and ZM2 models computed at the U.~S.~Naval Observatory and Pulkovo Observatory respectively, are full CPO models,
and can be considered providing the most complete correction to the a priori IAU precession-nutation model.
Different from that the SL model computed at the Paris Observatory is an FCN model.
The latter is recommended by the IERS Conventions (2010) for applications requiring highly accurate coordinate transformation.

Based on the test processing of the Intensives data collected in the framework of three IVS observing programs, Int1, Int2, and Int3,
using different CPO models we can draw several conclusions.

First, we numerically confirmed that proper CPO modelling is necessary to obtain highly accurate UT1 estimates from the Intensives.
Neglecting CPO leads to large systematic errors with amplitudes of up to 20~$\mu$as.
It was also demonstrated that an FCN model (e.g. SL) provides only partial compensation of this error as can be seen from Fig.~\ref{fig:dif} and
Table~\ref{tab:annual_term}.
A simple way to mitigate the difference between the IERS recommended SL model and CPO determined by VLBI (see Fig.~\ref{fig:cpo_dif}) would be
inclusion of the bias in the SL model, which has been intentionally excluded so far.
Although the SL model yet seems to be excessively smoothed, which may be important for other applications.

Our comparison of UT1 time series computed with NEOS, SL, and ZM2 models has shown practically no influence of CPO modelling on the
UT1 uncertainty, and the relative bias.
However, significant differences was found between these series including an annual term and (pseudo)-noise component.
It is important that even though no significant biases average over the whole time interval was found, the annual term can cause local (in time) biases
affecting operational combination.

For non-operational processing of the many-year Intensives time series, important for densification of the UT1 series obtained from the 24-hour VLBI
sessions, both NEOS and ZM2 models can be used with relatively small systematic differences between them, except some intervals where the difference
between the two models is substantial (see Figs.~\ref{fig:cpo_dif} and \ref{fig:dif}), e.g. in the end of 2008.
The systematic part of this difference can be presumably explained by the difference between the IVS and NEOS combined CPO series
(Fig.~\ref{fig:cpo_ivs}) which would be worth a special study.
However, the NEOS model seems to need stronger smoothing to remove the high-frequency signal coming most probably from incoming VLBI EOP series.
For real-time data processing, when one has to use predicted CPO, the ZM2 model may have an advantage due to better prediction accuracy (Malkin 2010).

Finally, it was shown that using different CPO models leads to significant, mostly systematic, differences in the UT1 estimates obtained
from the Intensives observing programs.
So, the best way to account for this effect, and corresponding recommendations to space geodesy analysis centers should be thoroughly discussed,
and a corresponding update of the IERS Conventions should be considered.
In particular, it seems to be advisable to use a CPO model which is the best compared to the IVS combined EOP series.
Also, a careful comparison of the IVS and IERS combined CPO series is needed in the framework of this discussion.

It should be noted that the obtained results and conclusions are applicable to the SLR and GNSS data processing as well,
in particular for precise orbit determination.

\section{Acknowledgements}
This paper is based on processing of VLBI observations collected on the international IVS network\footnote{ftp://cddis.gsfc.nasa.gov/vlbi/ivsdata/}.
The author wishes to express his gratitude to all the people working hard to make this data available.
Critical comments and valuable suggestions of three anonymous referees are highly appreciated.

\end{document}